\documentclass[]{mn2e}
\usepackage[dvips]{graphicx}

\title{Molecules in Southern Dense Cores I: Mopra and SEST Millimetre-Wave 
Observations}

\author[M. R. Hunt Cunningham et al.]
{Maria R. Hunt Cunningham,$^{1}$ John B. Whiteoak,$^2$ 
Paul A. Jones$^2$ and 
\newauthor 
Graeme L. White$^3$ \\
$^1$School of Physics, University of New South Wales, Sydney NSW 2052, Australia \\
$^2$Australia Telescope National Facility, PO Box 76, Epping NSW 1710,
Australia \\
$^3$ James Cook University, Townsville QLD 4811, Australia}

\date{Accepted 2003. Received 2003}
\pagerange{\pageref{firstpage}--\pageref{lastpage}}
\pubyear{2003}

\begin{document}

\label{firstpage}

\maketitle

\begin{abstract}
We have observed molecular-line transitions between 84 and 147 GHz in 24 dense southern molecular clouds with declinations south of -30 degrees using the Mopra and SEST radio telescopes. The results of observations of 2-mm and 3-mm transitions of the molecules CO, CS, C$_{2}$H, CH$_{3}$OH, HCN, HNC, HCO$^{+}$, HC$_{3}$N, OCS, HNCO and SO and several of their isotopomers are discussed in this paper, including ratios of the emission from the $^{12}$C, $^{13}$C and $^{18}$O isotopomers of CO, CS, HCO$^{+}$, HCN and HNC. This paper lists the calibrated observational data for each molecular cloud and discusses some general properties of this sample. The calibration process for the Mopra antenna of the ATCA is described in an appendix.\\
\end{abstract}

\begin{keywords}
ISM:molecules - radio lines:ISM - ISM:clouds - surveys
\end{keywords}

\section{Introduction}

Molecular-line emission provides a powerful method for investigating the events associated with star formation. Molecular emission can be used to find the temperature, density and kinematics of gas from which stars form, while emission from different molecules traces gas of differing physical conditions. The theoretical and observational underpinning for this type of study has been the work of many investigators over a thirty year period; a good review can be found in van Dishoeck \& Blake (1998).

The southern part of the Galactic plane includes the innermost parts of the galaxy as well as major dense molecular clouds and star-forming regions, many of which are difficult to observe with northern hemisphere telescopes due to their southern declinations. Consequently molecular line emission, molecular abundances and physical conditions have not been extensively investigated in clouds at declinations south of -30 degrees.

In the radio spectrum, molecular lines occur primarily at millimetre and submillimetre wavelengths. The commissioning of a millimetre wave receiver on the Mopra 
\footnote{The Mopra telescope is part of the Australia Telescope which is funded by the Commonwealth of Australia for operation as a National Facility managed by CSIRO.}
antenna of the Australia Telescope in 1994 provided the opportunity to observe molecular transitions at 3 mm with high velocity resolution, making it possible to produce a statistical database of molecular spectral-lines and molecular abundances in southern molecular clouds. 

In the study reported here, a sample of 24 molecular clouds with declinations south of -30 degrees has been observed with the Mopra Telescope in the 3-mm transitions of the molecules CO, CS, C$_{2}$H, CH$_{3}$OH, HCN, HNC, HCO$^{+}$, HC$_{3}$N, OCS, HNCO and SO and several of their isotopomers. In addition, complementary observations of 2-mm and 3-mm transitions have been obtained with the Swedish-ESO Submillimetre Telescope (SEST) 
\footnote{The SEST is operated jointly by the Swedish Natural Science Research Council and the European Southern Observatory.}

\section{ Observations and Equipment }
\subsection{Equipment}
\subsubsection{ The Mopra Telescope}

The 22-m `Mopra' antenna of the Australia Telescope (operated by the Australia Telescope National Facility, CSIRO and the University of New south Wales) is situated near Coonabarabran in New South Wales, Australia. The telescope is at latitude 31 degrees south and has an elevation of 850 metres above sea level. In 1994 the telescope was fitted with a millimetre-wave receiver allowing it to operate at frequencies between 85 and 116 GHz. The antenna has a `shaped' parabolic surface area to increase its forward gain and until 1999 it had a solid reflecting surface extending over a diameter of only 15 metres. Thus at millimetre wavelengths the reflecting surface had an effective diameter of 15 m, resulting in a half-power-beam-width that varied from 47 arcsec to 35 arcsec across the range 85 - 116 GHz. The antenna was resurfaced in 1999,
\footnote{ Resulting from funding contributions from the University of New South Wales, the solid surface was extended over the full diameter in 1999, extending the millimetre-wave collecting area to 22 metres.}
extending the effective millimetre-wave diameter to 22 metres, giving a half-power beamwidth varying from 35 to 24 arcsec over the same range of frequencies.

The Mopra mm-wave receiving system includes a cryogenically-cooled SIS (Superconductor Insulator Superconductor) amplifier that can be tuned to either single or double side-band mode. For single sideband operation, the response of the unwanted sideband can be decreased to at least 10 dB below the other sideband response using a calibration signal injected into both sidebands. A polarization splitter is used to separate the incoming signal into two channels that can be tuned separately, allowing the observation of two frequencies simultaneously. The `backend' is a digital autocorrelator, providing two simultaneous outputs with available band widths ranging from 4 to 256 MHz, which can be split into a maximum of 4096 channels. For the duration of this survey, the only observing method available was position switching: i.e. the observations were made by switching between the on-source position and a reference position with insignificant emission.

Both the antenna and the digital correlator have features in common to those used in the Australia Telescope Compact Array, and further details can be found in the Special Edition of the Journal for Electrical and Electronic Engineering, Australia, Vol 12, No 2 (1992). Technical details specific to the Mopra antenna can be found at www.narrabri.atnf.csiro.au/mopra/mopragu.pdf.

An important aspect of the study reported here (involving observations collected between 1994 and 2000) was the formulation of an intensity calibration scheme for the Mopra telescope. This calibration scheme is discussed in the appendix.

\subsubsection{ The SEST Telescope }

The SEST is situated at the European Southern Observatory at La Silla in Chile, about 600 km north of Santiago. The SEST has a Cassegrain design and a parabolic collecting surface with a diameter of 15 metres, giving a half power beam width ranging from 57 arcsec at 86 GHz to 15 arcsec at 346 GHz. 

The SEST is fitted with three SIS receivers that collectively cover the frequencies 78 - 117 GHz, 128 - 170 GHz, 215 - 270 GHz and 320 - 363 GHz. A polarization splitter allows the observation of two different frequencies simultaneously, although the combinations of frequencies are limited to 3-mm and 2-mm, or 3-mm and 1.6-mm.

Three Acousto-Optical Spectrometers (AOS) provide backends to the receivers. The high resolution AOS has a maximum bandwidth of 86 MHz, which can be divided into up to 2000 channels.

For more instrumentation details see Booth et al. (1989) or The SEST Handbook (1997).

\subsection{Selection of the Survey Sample }

The aim of the selection process was to select around 25 southern molecular clouds with strong molecular line emission and evidence of complex chemistry for detailed study, and involved a two-step process. Firstly, previous molecular-line surveys of the southern Galactic plane (Whiteoak \& Gardner 1974, Whiteoak \& Gardner 1978, Gardner \& Whiteoak 1978, Batchelor et al. 1981, Dickinson et al. 1982, Gardner \& Whiteoak 1984, Peters et al. 1986) were used to choose 50 candidate molecular clouds south of declination -30 degrees. Secondly, these candidates were observed with the Mopra radio telescope between September and November 1994, in the 3-mm transitions of CS, HCN and HCO$^{+}$ at frequencies of 98.0, 88.6 and 89.2 GHz respectively. These transitions were selected firstly because these molecules are generally abundant in molecular clouds and secondly because they were within the limited range of frequencies available on the Mopra telescope at that time. These observations were used to locate the line emission peaks, and select 24 of the brighter clouds for detailed study. The sample included several (5) dark clouds and also some objects showing unusual intensity ratios of the three transitions. Note that these observations were made during commissioning of the receiver, and before the development of better positional calibration and intensity calibration. Consequently, later observations of these transitions were obtained for use in the quantitative study (see table 3).

The list of 24 Galactic molecular clouds selected is given in table 1, with the observed positions in both B1950 and J2000 coordinates. The two molecular clouds Orion MC and M 17, which are near the celestial equator and well studied by radio telescopes in the northern hemisphere, have been included for comparison and calibration. Four separate pointing positions were chosen within the complex NGC 6334, and two positions within G291.3-0.7 (NGC 3576, RCW 57). Therefore, including the two reference clouds (Orion and M 17), there are 30 target positions in table 1.

The exact positions were chosen based on $^{13}$CO 1-0 observations at 110 GHz with the Mopra telescope in September 1995. The clouds were mapped over a region of a few square arcminutes to determine the position of the peak, and the angular size of the $^{13}$CO emission. The beam size was 37 arcsec, and the pointing errors 10-15 arcsec, with occasional discrepancies of up to 20 arcsec; the positions listed in table 1 are believed to be within 20 arcsec of the $^{13}$CO condensation peak. Later observations centred on these positions (table 3) have pointing errors better than 10 arcsec.

\begin{table*}
\caption{Galactic molecular cloud positions observed.}
\vspace{4mm}
\begin{tabular}{llcccc}
\hline
General    & Other   & Right Asc. & Decl. & Right Asc. & Decl. \\
Designation & Designation    & (J2000) & (J2000) & (B1950) & (B1950) \\ \hline
Orion M.C. & Orion KL & 05 35 14.5 & -05 22 29 & 05 32 47.0 & -05 24 22 \\
HH46 D.C.  &          & 08 25 37.8 & -51 04 00 & 08 24 10.6 & -50 54 08 \\
G265.1+1.5 & RCW 36   & 08 59 26.4 & -43 45 08 & 08 57 38.0 & -43 33 24 \\
G268.4-0.8 &          & 09 01 51.6 & -47 44 07 & 09 00 09.4 & -47 32 15 \\
Cham D.C.  &          & 11 06 32.6 & -77 23 35 & 11 05 08.1 & -77 07 20 \\
G291.3-0.7 & RCW 57 (E) & 11 12 04.1 & -61 18 29 & 11 09 56.1 & -61 02 09 \\
G291.3-0.7 & RCW 57 (W) & 11 11 47.4 & -61 19 28 & 11 09 39.6 & -61 03 09 \\
Coalsack D.C. &       & 12 01 27.7 & -65 07 54 & 11 58 53.8 & -64 51 11 \\
G301.0+1.2 & RCW 65   & 12 34 57.8 & -61 39 56 & 12 32 06.1 & -61 23 24 \\
G305.4+0.2 &          & 13 12 33.8 & -62 33 49 & 13 09 21.0 & -62 17 54 \\
G311.6+0.3 &          & 14 04 56.0 & -60 20 21 & 14 01 20.0 & -61 06 00 \\
G322.2+0.6 & RCW 92   & 15 18 38.0 & -56 39 00 & 15 14 47.6 & -56 28 05 \\
Lupus D.C. &          & 15 43 02.2 & -34 09 06 & 15 39 51.2 & -33 59 36 \\
G326.7+0.6 &          & 15 44 44.8 & -54 06 25 & 15 40 55.0 & -53 57 00 \\
G327.3-0.5 &          & 15 53 06.1 & -54 35 31 & 15 49 13.0 & -54 26 37 \\
G331.5-0.1 &          & 16 12 12.3 & -51 29 02 & 16 08 24.2 & -51 21 20 \\
G333.0-0.6 &          & 16 21 05.5 & -50 35 25 & 16 17 18.3 & -50 28 18 \\
G333.4-0.4 &          & 16 21 32.9 & -50 26 32 & 16 17 46.0 & -50 19 27 \\
G333.6-0.2 &          & 16 22 12.1 & -50 05 56 & 16 18 26.0 & -49 58 54 \\
G345.5+1.5 &          & 16 59 41.9 & -40 03 13 & 16 56 14.1 & -39 58 45 \\
G345.5+0.3 &          & 17 04 28.5 & -40 46 02 & 17 00 59.0 & -40 41 54 \\
NGC 6334 & NGC 6334 (S) & 17 19 55.9 & -35 57 53 & 17 16 34.5 & -35 54 51 \\
NGC 6334 & NGC 6334 (CO) & 17 20 23.4 & -35 55 00 & 17 17 02.0 & -35 52 00 \\
NGC 6334 & NGC 6334 (N) & 17 20 48.5 & -35 46 33 & 17 17 27.4 & -35 43 35 \\
NGC 6334 & NGC 6334 (N1) & 17 20 53.4 & -35 45 32 & 17 17 32.3 & -35 42 35 \\
G348.7-1.0 & RCW 122  & 17 20 05.2 & -38 57 22 & 17 16 38.3 & -38 54 21 \\
G351.6-1.3 &          & 17 29 12.8 & -36 40 12 & 17 25 49.8 & -36 37 50 \\
G353.4-0.4 &          & 17 30 24.6 & -34 41 39 & 17 27 05.0 & -34 39 23 \\
Cor Aust D.C. & R CrA & 19 10 18.2 & -37 08 58 & 19 06 56.0 & -37 13 54 \\
M 17       & M 17 SW  & 18 20 23.1 & -16 11 36 & 18 17 30.0 & -16 12 59 \\
\hline
\end{tabular}
\label{tab1}
\end{table*}

\subsection{Survey Observations}
\subsubsection{ Mopra Observations }

Molecular transitions between 85 and 116 GHz were observed with the Mopra telescope between September 1994 and September 1996 with the 15-metre solid millimetre surface, and in March and September 2000, after the resurfacing to 22 metres. Emission from 11 different molecules was observed (CO, CH$_{3}$OH, C$_{2}$H, CS, HCN, HC$_{3}$N, HCO$^{+}$, HNC, HNCO, OCS and SO), generally in more than one transition. Observations of more rare, isotopically substituted species of the molecules CO, CS, HCN, HCO$^{+}$ and HNC were also obtained ($^{13}$C for $^{12}$C and $^{18}$O for $^{16}$O). The actual transitions observed are listed in table 2, together with the date of observation, the frequency and the telescope used.

The single-sideband receiver temperature varied between 85 and 130 K. All observations were obtained with a bandwidth of 64 MHz divided into 1024 channels, giving a velocity resolution between 0.44 and 0.32 km s$^{-1}$ after Hanning smoothing. 

Observations were made in position switching mode, generally with a separation of 30 arcmin; however, for some extended sources such as the dark clouds, 60-arcmin separations were used. Periodic observations of SiO masers of known position were used to correct the pointing model, giving a pointing accuracy of better than 10 arcsec for observation obtained in 1995 or later. This includes all observations reported in table 3. 

\subsubsection{ SEST Observations }

The SEST was used in June 1999 to observe transitions between 84 and 147 GHz. The system temperatures were 170 - 250 K at 3 mm and 230-400 K at 2 mm wavelength bands. The beamwidth varied from 59 arcsec at 85 GHz to 35 arcsec at 147 GHz. Note that although the effective diameter of SEST was the same as Mopra (15 m) the Mopra dish was 'shaped' with different illumination resulting in a smaller beamwidth for a given frequency. A bandwidth of 42 MHz was used, divided into 1000 channels, giving velocity resolution between 0.30 and 0.17 km s$^{-1}$ after Hanning smoothing.

The observations were made using two different observing modes. Generally, dual-beam switching mode was utilised with a throw of 11 arcmin between the on-source and reference positions. However, for extended sources, where line emission was likely to be present in the reference position, frequency switching was used, with a frequency throw of 4 MHz. Periodic observations of SiO masers were used to check focus and pointing, giving a pointing accuracy better than 5 arcsec. Intensity calibration of the observations was performed on-line by the chopper wheel method (see Appendix), with the spectra obtained being on the T$_{A}$* scale.

\begin{table*} 
\centering
\caption{Molecular transitions observed}
  \begin{tabular}{llrll}
\hline
Molecule    & Transition     & Rest Frequency & Telescope & Date     \\
            &                &    (GHz)       &           & Observed \\    
\hline
CO          & 1-0            & 115.27120      & Mopra     & 1996 May \\
$^{13}$CO   & 1-0            & 110.20135      & Mopra     & 1996 May \\
C$^{18}$O   & 1-0            & 109.78216      & Mopra     & 1996 May \\
CH$_3$OH    & 5(-1)-4(0) E   &  84.52121      & SEST      & 1999 Jun \\
CH$_3$OH    & 15(3)-14(4) A- &  88.94009      & SEST      & 1999 Jun \\
CH$_3$OH    & 2(-1)-1(-1) E  &  96.73939      & Mopra     & 1995 Nov \\
CH$_3$OH    & 2(0)-1(0) A+   &  96.74142      & Mopra     & 1995 Nov \\
CH$_3$OH    & 2(0)-1(0) E    &  96.74458      & Mopra     & 1995 Nov \\
CH$_3$OH    & 2(1)-1(1) E    &  96.75551      & Mopra     & 1995 Nov \\
CH$_3$OH    & 3(1)-4(0) A+   & 107.01385      & SEST      & 1999 Jun \\
CH$_3$OH    & 3(0)-2(0) E    & 145.09375      & SEST      & 1999 Jun \\
CH$_3$OH    & 3(-1)-2(-1) E  & 145.09747      & SEST      & 1999 Jun \\
CH$_3$OH    & 3(0)-2(0) A+   & 145.10323      & SEST      & 1999 Jun \\
C$_2$H      & 1-0 3/2-1/2 F=1-1 &  87.28415   & Mopra     & 1995 Nov \\
C$_2$H      & 1-0 3/2-1/2 F=2-1 &  87.31692   & Mopra     & 1995 Nov \\
C$_2$H      & 1-0 3/2-1/2 F=1-0 &  87.32862   & Mopra     & 1995 Nov \\
CS          & 2-1            &  97.98096      & Mopra     & 1995 Nov \\
            &                &                & SEST      & 1999 Jun \\
$^{13}$CS   & 2-1            &  92.49429      & Mopra 22m & 2000 Mar \\
CS          & 3-2            & 146.96904      & SEST      & 1999 Jun \\
HCN         & 1-0 F=1-1      &  88.63041      & Mopra     & 1995 Oct \\
HCN         & 1-0 F=2-1      &  88.63184      & Mopra     & 1995 Oct \\
HCN         & 1-0 F=0-1      &  88.63393      & Mopra     & 1995 Oct \\
HC$_3$N     & 10-9           &  90.97899      & Mopra     & 1995 Nov \\
            &                &                & SEST      & 1999 Jun \\
HC$_3$N     & 11-10          & 100.07638      & SEST      & 1999 Jun \\
HC$_3$N     & 12-11          & 109.17363      & Mopra     & 1996 May \\
HC$_3$N     & 15-14          & 136.46440      & SEST      & 1999 Jun \\
HC$_3$N     & 16-15          & 145.56094      & SEST      & 1999 Jun \\
HCO$^+$     & 1-0            &  89.18851      & Mopra     & 1995 Oct \\
H$^{13}$CO$^+$ & 1-0         &  86.75429      & Mopra 22m & 2000 Oct \\
HC$^{18}$O$^+$ & 1-0         &  85.16225      & Mopra     & 1996 Sep \\
HNC         & 1-0 F=0-1      &  90.66345      & Mopra     & 1995 Nov \\
HNC         & 1-0 F=2-1      &  90.66357      & Mopra     & 1995 Nov \\
HNC         & 1-0 F=1-1      &  90.66365      & Mopra     & 1995 Nov \\
HN$^{13}$C  & 1-0            &  87.09086      & Mopra 22m & 2000 Oct \\
HNCO        & 4(2,3)-3(2,2)  &  87.89841      & Mopra     & 1996 Sep \\
HNCO        & 4(2,2)-3(2,1)  &  87.89862      & Mopra     & 1996 Sep \\
HNCO        & 4(0,4)-3(0,3)  &  87.92523      & Mopra     & 1996 Sep \\
OCS         & 7-6            &  85.13910      & Mopra     & 1996 Sep \\
OCS         & 8-7            &  97.30120      & Mopra     & 1996 Sep \\
OCS         & 9-8            & 109.46306      & Mopra     & 1996 Sep \\
OCS         & 12-11          & 145.94681      & SEST      & 1999 Jun \\
SO          & 3(2)-2(1)      &  99.29987      & Mopra     & 1995 Nov \\
\hline
\end{tabular}
\label{tab2}
\end{table*}


\subsection{Data Reduction}

All spectra were reduced using the CLASS (Continuum and Line Analysis Single-dish Software) package of the GILDAS (Grenoble Image and Line Data Analysis Software) working group.

The spectra were Hanning-smoothed to increase the signal-to-noise ratio before baseline subtraction and Gaussian fitting to features. The spectral profiles generally had complex shapes, so one or more Gaussian components were fitted. 

For Mopra spectra, the peak and integrated intensities of the Gaussian components were then corrected for atmospheric attenuation and scaled to the SEST corrected antenna temperature ($T_A^*$) scale (see Appendix for more details on intensity calibration of Mopra). The SEST corrected antenna temperatures ($T_A^*$) were then converted to main-beam brightness temperatures ($T_{MB}$) using the SEST main beam efficiency ($B_{eff}/F_{eff}$) which varied from 0.75 to 0.66 between 87 and 147 GHz (Booth et al., 1989, SEST Observers Handbook). 

\section{Results}

Table 3 lists the observed molecular transitions, and the results of the Gaussian-component fits for each molecular cloud. An excerpt from table 3, showing results for the molecular cloud G265.5+1.5, is given in the body of this paper. The complete table can be found in the electronic version of the paper and on the VizieR data base of astronomical catalogues at the Centre de Donn$\acute{e}$es astronomiques de Strasbourg (CDS) website. The first two columns list the molecule and the transition observed. The next three columns are the Gaussian fit parameters (central velocity, full-width at half maximum and peak expressed as main beam brightness temperature T$_{MB}$). Multiple Gaussians were used to fit complex profiles. However, interpreting components in terms of separate clouds is difficult because in some cases, double-peaked profiles fitted as two Gaussians may be a broad emission line with a self-absorption dip. The velocities of the HCN hyperfine components are for the J = 2-1 component with the F = 1-0 and F = 1-1 line centres fitted using the same velocity scale. These latter hyperfine transitions are shown in brackets in the table. (This is because the components are often blended, so it was not feasible to velocity correct the hyperfine components separately.) The 6th column shows the integrated line emission ($\int$ T$_{MB}$ dv) for individual Gaussian components, while the 7th column shows the total line emission in the transition, obtained by adding the individual Gaussian components; in the case of HCN this total includes the emission in the three hyperfine components. The main beam brightness temperature (T$_{MB}$), is the true brightness temperature of the source, if the source just fills the main beam. If the source angular size is known, then this temperature can be deconvolved with the source size to give the actual source brightness temperature.

The 8th column gives the uncertainty in the integrated brightness temperature (column 7). For Mopra observations, the scatter of repeated observations of a source was used to make an estimate of the accuracy of the integrated brightness temperature of the lines. The error in integrated brightness was considered to be made up of two parts adding in quadrature - a constant percentage error B, which dominates at high brightness, and a constant absolute error A, due to noise in the spectra, which dominates at low brightness. From the scatter of repeated observations of strong lines ($\int$T$_{MB}$ dv $ >$ 5 K km s$^{-1}$) we determine B = 12 \%, and from repeated observations of weak lines ($\int$T$_{MB}$ dv $<$ 5 K km s$^{-1}$) we determine A = 0.24 K km s$^{-1}$. For a typical linewidth of 4 km s$^{-1}$, this value of A corresponds to peak brightness 0.06 K and is consistent with the contribution to line fitting from the noise in the spectra. 

The calibration uncertainties of the SEST observations are assumed to be 10 \% at 3 mm and 15 \% at 2 mm (SEST Observers Manual 2003; Heikkila 1998), and the total uncertainty the sum in quadrature of the calibration uncertainty and the formal errors in fitting the lines using the CLASS fitting routine (GAUSS). The Mopra and SEST uncertainties turn out to be similar (listed in column 8 of table 3).

\begin{table*}
\centering
\caption{ Fitted line intensities from Mopra and SEST observations for the molecular cloud G265.5+1.5 (RCW 36). The remainder of the table appears in the electronic version of this paper.}
\begin{tabular}{llrrrrrrrrr}
\hline
&& \multicolumn{4}{c}{ \textbf{Gaussian Components of fit}} && \multicolumn{2}{c}{ \textbf{Total Emission} }\\
Molecule&Transition& Centre&dv&T$_{MB}$&$\int$T$_{MB}$dv&&$\int$T$_{MB}$dv&$\delta\int$T$_{MB}$dv\\
& &(km s$^{-1}$)&(km s$^{-1}$)&(K)&(K km s$^{-1}$)& &(K km s$^{-1}$)&(K km s$^{-1}$)\\
\hline
$^{12}$CO&1-0&1.4&1.3&14.25&19.73&&19.7&2.3\\
$^{13}$CO&1-0&4.0&0.9&6.90&6.62&&6.6&0.8\\
C$^{18}$O&1-0&4.5&0.6&1.08&0.69&&0.7&0.3\\
&&&&&&&&\\
C$_{2}$H&1-0  3/2-1/2  F=1-0&4.4&2.0&0.97&0.25&&0.7&0.3\\
&1-0  3/2-1/2  F=2-1&1.1&0.8&0.39&0.43&&&\\
&&&&&&&&\\
CS&2-1&5.3&0.9&1.04&0.99&&1.0&0.3\\
CS&3-2&5.2&0.7&0.56&0.45&&0.4&0.2\\
&&&&&&&&\\
HCN&1-0, F=2-1&5.3&1.1&0.87&1.02&&1.8&0.3\\
&1-0, F=1-1&(10.1)&1.0&0.38&0.40&&&\\
&1-0, F=0-1&(-1.7)&1.0&0.33&0.35&&&\\
H$^{13}$CN&1-0&-&2.4&0.31&0.78&&0.8&0.2\\
&&&&&&&&\\
HNC&1-0&5.2&0.9&0.72&0.67&&0.7&0.3\\
HN$^{13}$C&1-0&&&-&&&$<$0.2&\\
&&&&&&&&\\
HNCO&4(0,4)-3(0,3)&&&-&&&$<$0.10&\\
&&&&&&&&\\
HCO$^{+}$&1-0&5.3&1.1&1.40&1.63&&1.6&0.3\\
H$^{13}$CO$^{+}$&1-0&4.9&1.0&0.39&0.41&&0.4&0.1\\
HC$^{18}$O$^{+}$&1-0&&&-&&&$<$0.14&\\
&&&&&&&&\\
HC$_{3}$N&10-9&&&-&&&$<$0.04&\\
HC$_{3}$N&11-10&&&-&&&$<$0.07&\\
HC$_{3}$N&12-11&&&-&&&$<$0.06&\\
HC$_{3}$N&15-14&&&-&&&$<$0.08&\\
HC$_{3}$N&16-15&&&-&&&$<$0.08&\\
&&&&&&&&\\
OCS&7-6&&&-&&&$<$0.14&\\
OCS&8-7&&&-&&&$<$0.09&\\
OCS&9-8&&&-&&&$<$0.23&\\
OCS&12-11&&&-&&&$<$0.09&\\
&&&&&&&&\\
CH$_{3}$OH&2(0)-1(0) A+&&&-&&&$<$0.31&\\
CH$_{3}$OH&3(-1)-2(-1)E&&&-&&&$<$0.13&\\
CH$_{3}$OH&15(3)-14(4)A-&&&-&&&$<$0.04&\\
&&&&&&&&\\
SO&3(2)-2(1)&5.2&0.7&1.29&0.99&&1.0&0.3\\
\hline
\end{tabular}
\end{table*}


\begin{figure*} 
\begin{center}
\includegraphics[angle=-90,width=20.5cm]{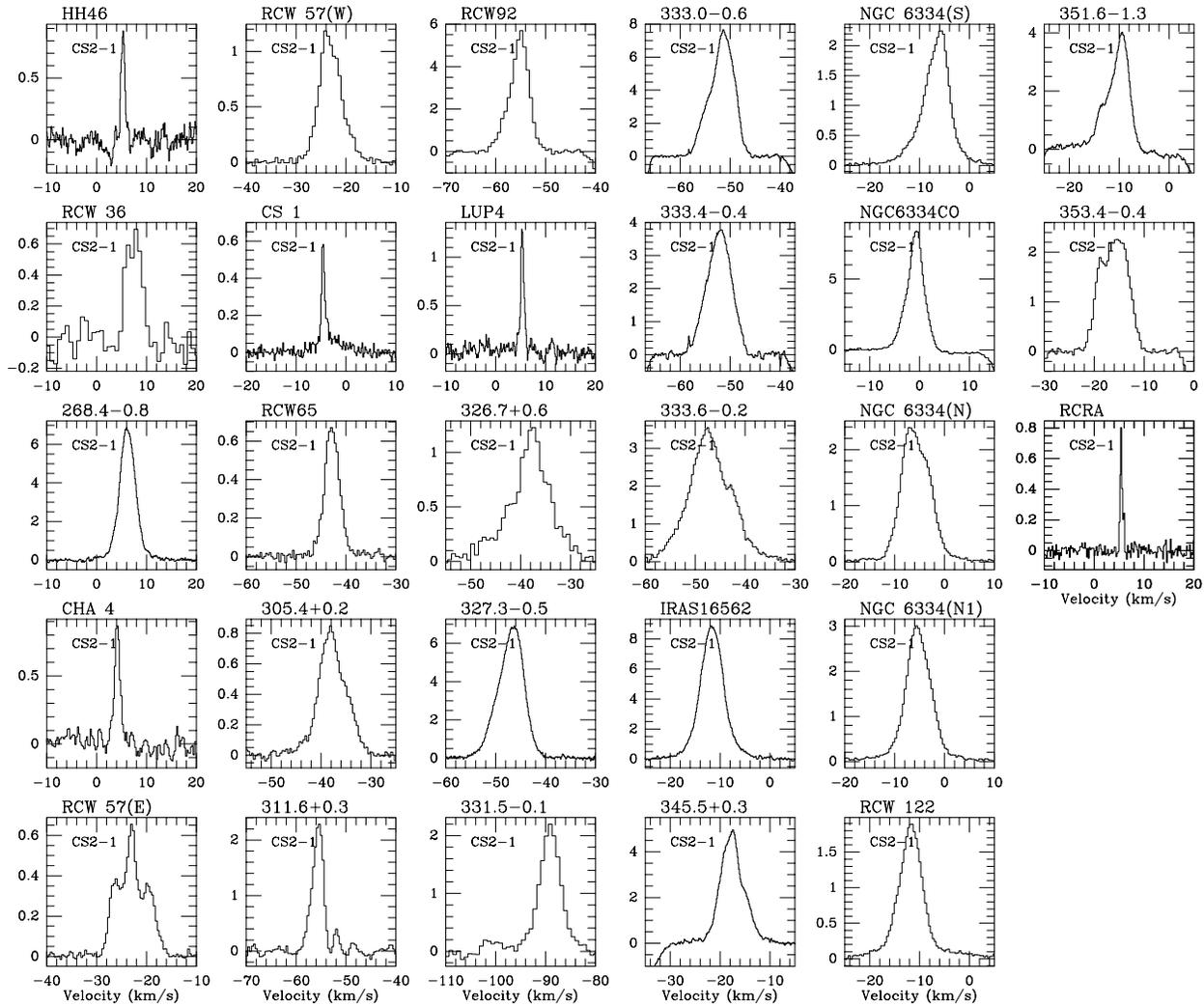}
\caption{CS 2-1 spectra towards each of the molecular clouds in table 1. This transition shows features characteristic of the many of the other molecular transitions observed. The name of each molecular cloud is shown above each spectrum.}
\label{spectra}            
\end{center}
\end{figure*}

\begin{figure*} 
\begin{center}
\includegraphics[angle=-90,width=23cm]{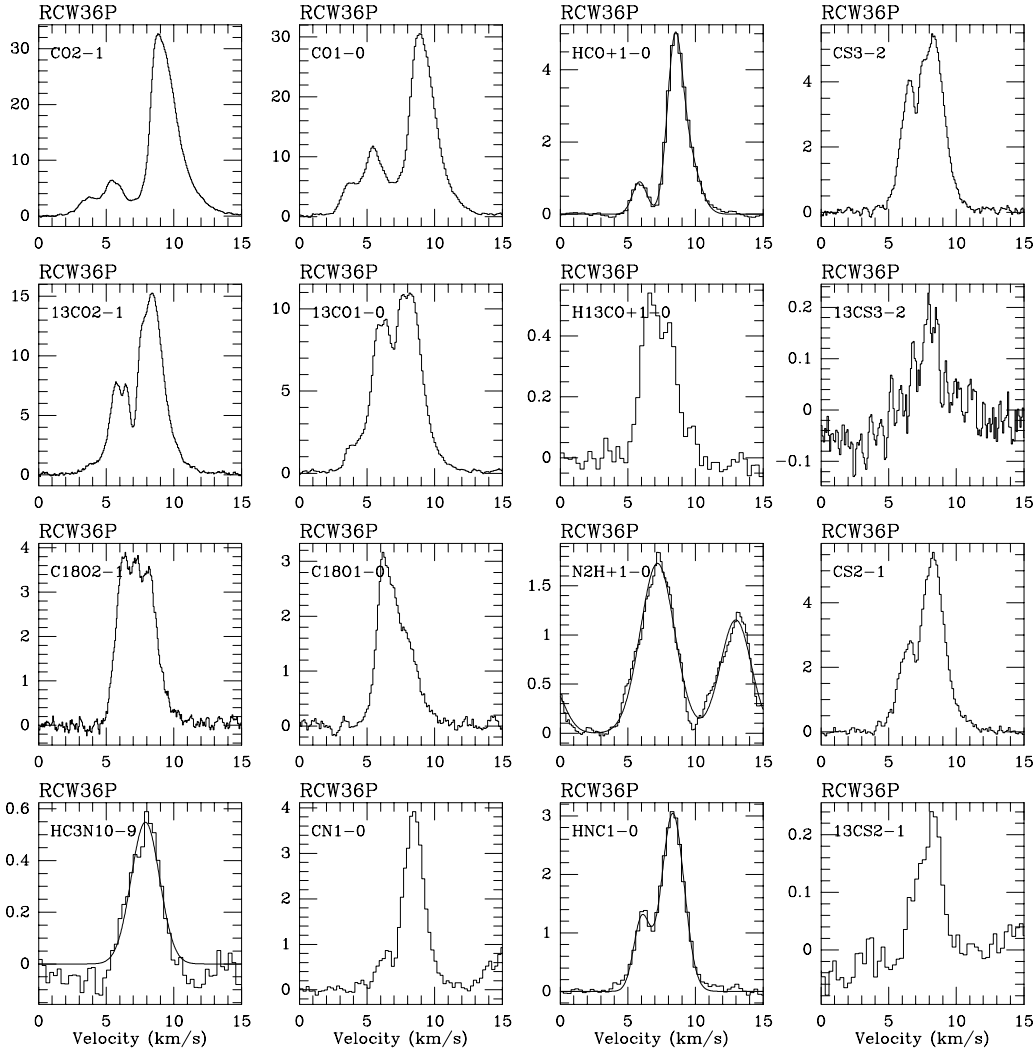}
\caption{Spectra towards the molecular cloud G265.5+1.5. The transition is shown in the top left-hand corner of each box. The differences between the line profiles of the different transitions are typical of many of the molecular clouds in the sample.}
\label{spectra2}            
\end{center}
\end{figure*}

\begin{figure*} 
\begin{center}
\includegraphics[angle=-90,width=14cm]{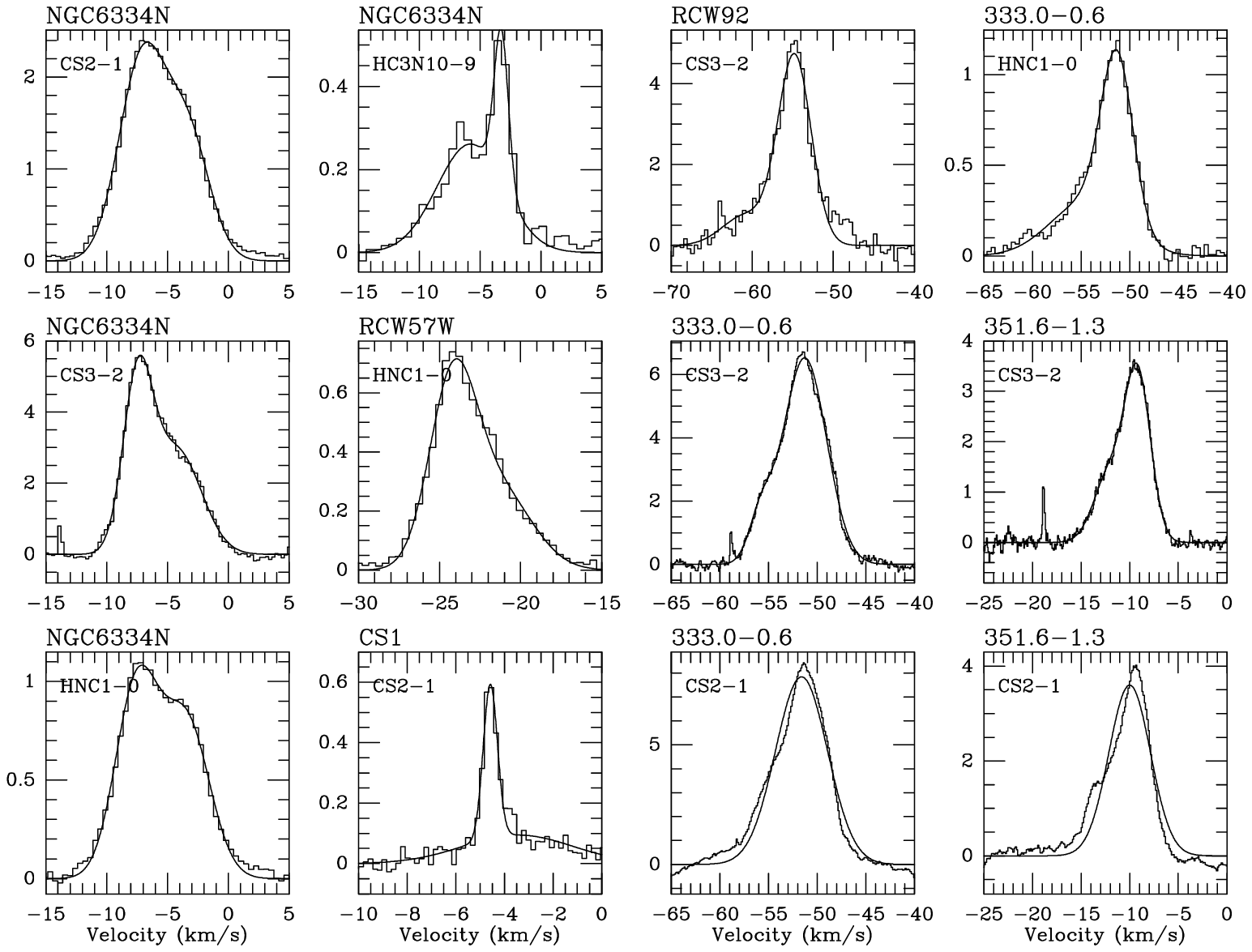}
\caption{Examples of spectra showing line wings and asymmetric line profiles, suggesting the presence of outflow.}
\label{spectra3}            
\end{center}
\end{figure*}

\section{Discussion}

Some general properties of the line emission and profiles are discussed in this section, including the ratios of the emission from different isotopomers of several molecules. In addition, the line intensities have been used to calculate molecular abundances using both LTE and LVG models, which we present in a separate paper (Hunt Cunningham et al. 2004). 

\subsection{Spectral Line Intensities and Profiles}

Figure 1 shows CS 2-1 spectra towards each position listed in table 1. Each spectrum is plotted over a velocity interval of 30 km s$^{-1}$. This transition was chosen to illustrate the characteristic line profile in each molecular cloud as it is ubiquitous, generally has good signal to noise, and tends to have a `typical' line profile. On the other hand, the CO spectra for each molecular cloud often show emission features covering a significantly wider range of velocities than found for other molecules, as can be seen in figure 2, which shows spectra for a number of different transitions towards the molecular cloud G265.1+1.5 (RCW 36). The spectra towards this molecular cloud illustrate features common to many of the molecular clouds. 

Self-absorption is evident in the $^{12}$CO spectra towards about a third of the molecular clouds, and can also be seen in some spectra of $^{13}$CO and C$^{18}$O. Transitions that are believed to be self-absorbed are marked in table 3 by an asterisk. Self-absorption is also evident in some clouds in transitions of CS, HCN, HCO$^{+}$ and SO, but not generally in transitions of OCS, HC$_{3}$N and CH$_{3}$OH, implying that these latter transitions are more likely to be optically thin. In addition to the presence of self-absorption, the spectra, particularly those of $^{12}$CO, often show a complex emission/velocity structure that cannot be adequately fitted by a single Gaussian. As expected, emission from the dark clouds in the sample (HH 46, the Coal Sack, the Chameleon, Lupus and Corona Australis) is distinguished by narrow line widths compared to other clouds in the sample (see figure 1).

High-velocity wings and asymmetric line profiles are seen in the spectra of a number of molecular clouds (e.g spectra of CO, CS, HCN and SO observed towards the positions NGC 6334(CO) and NGC 6334 (N)) suggesting that outflows, presumably from young stellar objects (YSOs), are present. Figure 3 shows some examples of spectra with line wings and asymmetries.

\subsection{Isotopomer Intensity Ratios}

The $^{12}$CO emission is found to be optically thick for all molecular clouds in this sample. Table 4 lists the emission intensity ratios for $^{12}$C:$^{13}$C isotopomers of CO, CS, HCN, HNC and HCO$^{+}$, as well as the sample standard deviation ($\sigma_{n-1}$), the minimum and maximum values of the ratio, and the number of molecular clouds for which the ratio has been calculated. The terrestrial abundance ratio of the atoms $^{12}$C:$^{13}$C is 89, the ratio of the abundances of the molecules $^{12}$CO:$^{13}$CO in nearby molecular clouds is found to be around 50; however, significant variations in this ratio have been found in individual molecular clouds, and it is likely that the ratio varies with galactocentric distance. Langer and Penzias (1990) have found a value of 57 for molecular clouds at the Sun's distance from the Galactic Centre, with values ranging from 25 near the centre of the Galaxy to 70 at a distance of 12 kpc. In contrast, the mean ratio for the intensity of $^{12}$CO emission (hereafter I$^{12}$) to emission from $^{13}$CO (hereafter I$^{13}$) in table 4 is 3.1 $ \pm $ 0.2 for this sample of southern molecular clouds, implying that the $^{12}$CO emission is optically thick for all molecular clouds of the sample. This is very close to the value for I$^{12}$: I$^{13}$ of 3.0 $ \pm $ 0.9 found for a sample of northern Galactic-plane GMCs between l = 34$^{o}$ and l = 51$^{o}$ (Polk et al. 1988) suggesting that there is no systematic difference in this ratio between the northern and southern Galactic plane. Doty \& Neufeld (1997) have modelled I$^{12}$: I$^{13}$ for the dense quiescent star-forming cores of molecular clouds, assuming an approximately terrestrial abundance ratio of 100:1 for $^{12}$CO:$^{13}$CO. They have predicted an I$^{12}$: I$^{13}$ ratio of between 2 and 6, which is consistent with the findings of this study, and again suggests that the $^{12}$CO emission has high optical depth for all clouds in this sample. 


\begin{table*}
\centering
\caption{Emission intensity ratios for $^{12}$C:$^{13}$C from isotopomers of 
CO, CS, HCN, HNC and HCO$^{+}$.}
\vspace{4mm}
\begin{tabular}{lrrrrr}
\hline \\ [-2mm]
   & {\large $\frac{CO}{^{13}CO}$} & {\large $\frac{CS}{^{13}CS}$} & 
 {\large $\frac{HCN}{H^{13}CN}$} & {\large $\frac{HNC}{HN^{13}C}$} & 
 {\large $\frac{HCO^{+}}{H^{13}CO^{+}}$} \\ [2mm]


\hline
 Mean           & 3.1  & 12.0 & 7.8  & 11.9 & 6.8  \\
 Standard dev.  & 1.9  & 5.7  & 5.4  & 5.6  & 7.6  \\
 Minimum        & 1.7  & 6.1  & 0.9  & 2.7  & 0.3  \\
 Maximum        & 10.0 & 30.2 & 25.1 & 21.7 & 33.3 \\
 Number of Obs. & 33   & 19   & 25   & 20   & 28   \\ 
\hline
\end{tabular}
\label{tab4}
\end{table*}


Table 5 lists the same values for the intensity ratio of $^{13}$C: $^{18}$O isotopomers. The mean ratio of I$^{13}$ to emission from C$^{18}$O (hereafter I$^{18}$) is 6.9 $ \pm $ 0.4. This is somewhat higher than the value for this ratio of 5.5 that would be expected for optically thin emission if terrestrial ratios for atomic $^{12}$C:$^{13}$C and $^{16}$O: $^{18}$O were assumed. The values for I$^{13}$: I$^{18}$ for individual clouds range from 2.5 to 15.2, so in some molecular clouds the ratio is far higher than expected from the terrestrial ratio of the isotopes. 

The ratio I$^{12}$: I$^{18}$ found in this study is 25 $ \pm $ 3, with a range of 4.2 to 98. Doty \& Neufield (1997), assuming terrestrial ratios, predicted that I$^{12}$: I$^{18}$ should be between 5 and 20, assuming that $^{12}$CO and C$^{18}$O were coextensive. A possible reason for the high ratios found in this study is that the emission from $^{12}$CO arises from a larger volume of gas than does emission from C$^{18}$O. This suggestion is supported by the average line-widths for $^{12}$CO and C$^{18}$O emission for the sample. The average line-width for $^{12}$CO is 8.7 km s$^{-1}$, while the average line-width for C$^{18}$O is 4.1 km s$^{-1}$, suggesting that $^{12}$CO emission arises from a larger volume of the gas (assuming that line-width is proportional to radius, as if will be if there is a linear velocity gradient across the molecular cloud). However, note that this difference in average line widths may be due to optical depth effects. If the $^{12}$CO emission is optically thin in the wings of the line profiles then the C$^{18}$O emission in the line wings will not be detected, assuming a terrestrial abundance of $^{12}$CO:C$^{18}$O of 490:1.


\begin{table}
\centering
\caption{Emission intensity ratios for $^{13}$CO:C$^{18}$O from isotopomers of CO and HCO$^{+}$.}
\vspace{4mm}
\begin{tabular}{lrr}
 \hline \\ [-2mm] 
   & {\large $\frac{ ^{13}CO }{ C^{18}O } $ }  &
     {\large $\frac{ H^{13}CO^+ }{ HC^{18}O^+ } $ }  \\ [2mm]
\hline
Mean           & 6.9  & 15.3 \\
Standard dev.  & 2.6  & 13.7 \\
Minimum        & 2.5  & 6.3  \\
Maximum        & 15.2 & 60.0 \\
Number of Obs. & 33   & 15   \\
\hline
\end{tabular}
\label{tab5}
\end{table}


The average line-width of $^{13}$CO is 5.3 km s$^{-1}$, similar to that for C$^{18}$O, suggesting that the C$^{18}$O and $^{13}$CO emission arise from a similar volume of gas. If this is the case, then the optical depth of the $^{12}$CO emission is likely to be even higher than that implied by the emission ratios in Table 4. The critical densities for $^{13}$CO and C$^{18}$O are lower than that for $^{12}$CO, implying that the difference in volume of the emitting gas is due to a lower abundance of the more rare isotopomers in outer regions of the molecular cloud rather than to varying excitation conditions. The unexpectedly high values of I$^{13}$:I$^{18}$ can be understood if the $^{13}$CO and C$^{18}$O emission are not completely co-extensive, with the C$^{18}$O emission arising from a smaller volume of gas. The relative average line widths (4.1 km s$^{-1}$ for C$^{18}$O, 5.3 km s$^{-1}$ for $^{13}$CO) provide some support for this assertion. It is possible that self-shielding (from dissociating UV radiation both within and incident on the molecular cloud) in the more abundant isotopomers accounts for their apparently wider distribution within the molecular gas (Langer \& Penzias 1990).

The average ratio of emission from H$^{13}$CO$^{+}$ to that from HC$^{18}$O$^{+}$ is 15.3 $ \pm $ 3.5 (Table 5). Once again, this is far higher than would be expected from optically thin emission from both isotopomers if terrestrial ratios of atomic $^{12}$C: $^{13}$C and $^{16}$O: $^{18}$O apply. The high ratio suggests that the HC$^{18}$O$^{+}$ emission may arise from a smaller volume of gas than the H$^{13}$CO$^{+}$ emission, once again possibly due to self-shielding in the more abundant isotopomer (H$^{13}$CO$^{+}$). The higher ratio for emission from H$^{13}$CO$^{+}$:HC$^{18}$O$^{+}$ compared with that from $^{13}$CO:C$^{18}$O may also support this suggestion. HC$^{18}$O$^{+}$ can be expected to be far less abundant than C$^{18}$O, and consequently full self-shielding from dissociating UV radiation will occur at a significantly smaller cloud radius. On the other hand, the high ratio may be due to chemical fractionation effects. Chemical fractionation (the preference for a molecule consisting of one isotope over another in a chemical reaction) can lead to different isotope ratios in the molecular products of a reaction with respect to the reactants. More complex products have a higher ratio of the lighter to the heavier isotope (e.g a higher $^{16}$O: $^{18}$O ratio) (Engel \& Macko 1997), so HC$^{18}$O$^{+}$ may be less abundant with respect to its $^{13}$C isotopomer than C$^{18}$O.

The average emission ratios of the $^{12}$C to $^{13}$C isotopomers of CS, HCN, HNC and HCO$^{+}$ (Table 4) suggest that the $^{12}$C isotopomer of these molecules is generally optically thick in dense molecular clouds, although less so than $^{12}$CO. A surprising result is that the minimum value for the emission ratio HCN:H$^{13}$CN is 0.9, and the minimum value for HCO$^{+}$:HC13O$^{+}$ is 0.3. Both these low values occur in the same molecular cloud, the Chameleon dark cloud. No other molecular clouds have values less than unity for these ratios. In fact, all ratios for emission from $^{12}$C isotopomers to $^{13}$C isotopomers in the Chameleon dark cloud are low compared to other molecular clouds. The reason for this is not clear. It is possible that the Chameleon dark cloud has a low elemental ratio of $^{12}$C to $^{13}$C, but further observations of the Chameleon dark cloud will be required to clarify the situation.

\section{Summary} 

Transitions of 11 different molecules and several of their isotopomers have 
been observed in a sample of 26 southern molecular clouds. The observational 
results are as follows:

1. The CO spectra often show a significantly wider range of velocities than 
found for other molecules. 

2. Self-absorption is present in about a third of the CO profiles, and in 
some profiles of $^{13}$CO, C$^{18}$O, CS, HCN, HCO$^+$ and SO.

3. Self-absorption is not generally observed in transitions of OCS, HC$_3$N 
and CH$_3$OH, suggesting that these transitions are optically thin in most 
molecular clouds of the sample, or that these transitions only exist in the 
higher-excitation, inner parts of the clouds.

4. CO is optically thick in all molecular clouds of the sample. The average 
emission ratio for $^{12}$CO:$^{13}$CO is $3.1 \pm 0.2$, very close to values 
found in surveys of northern hemisphere molecular clouds, and predicted by 
models.

5. The average emission ratio for $^{13}$CO:C$^{18}$O is $6.9 \pm 0.4$, 
higher than would be expected for optically thin emission if Solar System 
ratios for $^{13}$C:$^{18}$O were assumed. This may be because C$^{18}$O 
occupies a smaller volume of gas than does $^{13}$CO, due to self-shielding 
of $^{13}$CO.

6. The average emission ratio of H$^{13}$CO$^+$:HC$^{18}$O$^+$ is 
$15.3 \pm 3.5$, also higher than would be expected for optically thin 
emission if Solar System ratios for $^{13}$C:$^{18}$O were assumed. 
Self-shielding in H$^{13}$CO$^+$ is a possible explanation, however 
chemical fractionation effects may also be responsible for the high ratio.

7. CS, HCN, HNC and HCO$^+$ are generally found to be optically thick for 
the molecular clouds of the sample.

8. The Chameleon molecular cloud has $^{12}$C:$^{13}$C isotopomer ratios of 
less than unity for HCN and HCO$^+$, and has low $^{12}$C:$^{13}$C 
isotopomer ratios for all molecules. Further observations will be required 
to clarify the reasons for this situation. 

\section*{Acknowledgments}


This research forms part of the PhD thesis of MRHC undertaken at the 
University of Western Sydney.
MRHC acknowledges support from an Australian Postgraduate 
Award and thanks DIST Access to Major Research Facilities Scheme for travel support to SEST. 
MRHC also gratefully acknowledges support from the ATNF, including free accommodation and use of computing facilities for the duration of this work.
The Mopra telescope is part of the Australia Telescope which is funded
by the Commonwealth of Australia for operation as a National Facility
managed by CSIRO.

\section*{Appendix: Calibration of the Mopra Telescope}

The accurate intensity calibration of the observations reported here is important, 
as they have been used to calculate abundances (Hunt Cunningham, 
in preparation). At millimetre wavelengths there is significant emission and 
absorption of radiation by the atmosphere, and the component due to water 
vapour in the atmosphere can vary over timescales as short as a few minutes. 
One way to correct for these optical depth fluctuations is the `chopper wheel'
method, where an ambient temperature blackbody source is periodically 
inserted into the antenna beam as a reference (see e.g. Rohlfs \& Wilson 1996; 
Kutner \& Ulich 1981). This method as applied to both the Mopra telescope and 
the SEST is outlined here. More information on the terms and definitions used 
below can be found in Rohlfs \& Wilson (1996).

The antenna temperature directly measured at the telescope ($T_A$) needs to 
be corrected by various factors to account for atmospheric attenuation, 
aperture efficiency, far-sidelobe contributions (back scattered radiation) and 
electronic losses and noise (receiver temperature). A quantity known as the 
calibrated antenna temperature ($T_A'$) is corrected for atmospheric 
attenuation, backscattered radiation and receiver temperature, while what is 
known as the corrected antenna temperature ($T_A^*$) is also corrected for 
the forward efficiency of the telescope ($F_{eff}$), so 
$ T_A^* = T_A'/F_{eff}.$

The forward beam includes the sidelobes in the forward direction, so a better 
parameter to use is the main beam brightness temperature ($T_{MB}$), where \\
\[ T_{MB} = T_A'/B_{eff} = T_A^* F_{eff}/B_{eff}. \]

The SEST observations were made with on-line chopper wheel calibration system, 
and the observations were corrected by the on-line software to the corrected 
antenna temperature $T_A^*$, so only the (frequency dependent) correction 
$F_{eff}/B_{eff}$ needed to be applied to get $T_{MB}$.
 
The Mopra system did not have such on-line chopper wheel or software, so a 
`blackbody paddle' was inserted every few minutes to measure the gated total 
power (GTP) which could then be used off-line to calibrate the antenna 
temperature $T_A$ for atmospheric attenuation and instrumental parameters. 

Let $T_R$ be the receiver temperature, $T_S$ the source temperature, $T_a$
the ambient temperature of the atmosphere and $\tau$ the optical depth of the 
line of sight through the atmosphere ($\tau = \tau_0\sec(Z)$, 
where $Z$ is the zenith angle and $\tau_0$ the optical depth vertically 
through the atmosphere).
The on-source signal $S$ is then 
\[ S = C [T_R + T_a(1 - \exp(-\tau)) + T_S \exp(-\tau)], \]
where $C$ is a constant, and the off-source reference $R$ is
\[ R = C [T_R + T_a(1 - \exp(-\tau))]. \]

The usual position switching, subtracting the reference from the signal, 
gives  
\[ S - R = C T_S \exp(-\tau) \]
correcting for the receiver temperature and the atmosphere emission, but not 
for the atmospheric absorption of the source emission.

The paddle reference can be considered as a black body, at the ambient 
temperature $T_a$. The power $P_r$ at the off-source position is 
\[ P_r = C' [T_R + T_a(1 - \exp(-\tau))] \]
where $C'$ is another constant,
and the power $P_p$ with the black-body paddle in the telescope beam is
\[ P_p = C' [T_R + T_a]  \mbox{ so } P_p - P_r = C' T_a \exp(-\tau). \]

This can be used to correct the observations for atmospheric attenuation by
\[ (S - R)/(P_p - P_r) = (C/C')(T_S/T_a) \mbox{ or } \]
\[    T_S = (C'/C)T_a (S - R)/(P_p - P_r) \]
where $C$ and $C'$ are constants of the telescope hardware and software to 
be determined and calibrated out.

A complication of the Mopra spectral calibration here is that the observing 
program SPECTRA did not output the difference $S - R$ of the signal and 
reference spectra, but the quotient 
\[ Q = K [(S/R) - 1] T_R = K T_R (S - R)/R \]
where $K$ is a constant required from the instrumental calibration.

The standard calibration method equation is modified to
\[ Q P_r/(P_p - P_r) = K T_R (S - R) P_r / [R (P_p - P_r)] \]
\[ = K T_R T_S/T_a \mbox{  or } \] 
\[ T_S = Q (T_a/K T_R) P_r/(P_p - P_r). \]

The Mopra telescope was not calibrated absolutely, for the period that these 
observations were made, so the factor $K$ was significant, and we had to do a
relative calibration by comparing Mopra and SEST observations of the calibrator
sources Orion MC and M17. The integrated line intensities 
(in K km s$^{-1}$) were used for 43 lines between 86 and 115 GHz observed 
at Mopra several times over a 
range of weather conditions. Before calibration, the scatter of repeated 
observations in $Q$ about the mean for that line was rms 17 \% with some data 
points more than 30 \% from the mean, indicating the effect of variable 
atmospheric absorption. After calibration, the scatter about the mean was 
7 \%, showing the effectiveness of the correction for absorption. Figure 4 
shows the uncalibrated Mopra integrated antenna temperatures ($\int Q \,dv$)
plotted with the SEST integrated, corrected antenna temperatures 
($\int T_A^* \,dv$), while figure 5 shows the same data after the 
calibration procedure outlined above.

The Mopra spectra quotient $Q$ values were converted to the SEST corrected 
antenna temperature scale $T_A^*$ with a factor $1.62 \pm 0.06$ accounting 
for the instrumental factors in the Mopra SPECTRA $Q$ scale. 

\begin{figure} 
\begin{center}
\includegraphics[angle=-90,width=6.5cm]{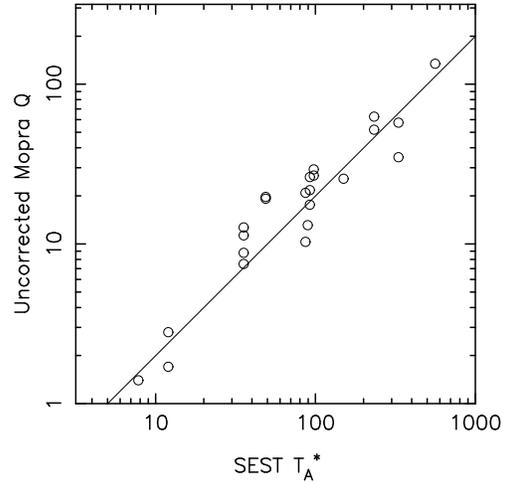}
\caption{ Uncorrected Mopra integrated brightness temperatures ($Q$ values) plotted with SEST calibrated integrated brightness temperatures. The uncalibrated Mopra values show a large scatter of around 40 \%, due to uncalibrated atmospheric attenuation, as well as a scale error due to unknown instrumental parameters when compared to the calibrated SEST data.}
\label{Mopra_cal_1}            
\end{center}
\end{figure}

\begin{figure} 
\begin{center}
\includegraphics[angle=-90,width=6.5cm]{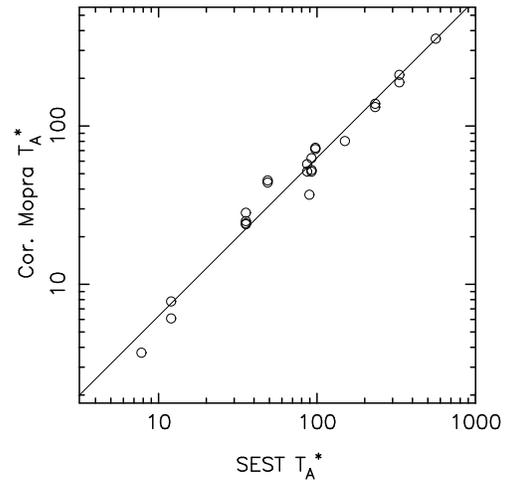}
\caption{ The corrected Mopra $T_A^*$ values, show a much tighter correlation with the SEST values (c.f. figure 4), with scatter around 20 \%, and the comparison enables the scale factor of 1.62 to be determined.}
\label{Mopra_cal}            
\end{center}
\end{figure}


\bsp 

\label{lastpage}
\end{document}